\documentclass[a4paper,11pt]{article}
\usepackage{setspace}
\usepackage[left = 2cm, right=2cm, top=3cm, bottom=3cm]{geometry}
\usepackage[pdftex]{graphicx}
\usepackage{cite}
\usepackage{bm}
\usepackage{float}
\usepackage{amsmath,amssymb,latexsym}
\usepackage{color}
\usepackage[T1]{fontenc}
\usepackage{wasysym}
\usepackage{siunitx}
\usepackage{setspace}
\usepackage{relsize}
\usepackage{tikz}
\usepackage{lineno, blindtext}
\usepackage{authblk}
\usepackage[symbol]{footmisc}
\usepackage{lineno}
\usepackage[pdfencoding=auto, psdextra]{hyperref}

\begin{document}
	
	
	
	\renewcommand{\figurename}{Fig.}
	
	\title{\color{blue}\textbf{Pattern selection in radial displacements of a confined aging viscoelastic fluid}}
	\author[1, $\dagger$]{Palak Palak}
	\affil[1]{\textit{Soft Condensed Matter Group, Raman Research Institute, C. V. Raman Avenue, Sadashivanagar, Bangalore 560 080, INDIA}}
	\author[1, $\ddagger$]{Vaibhav Raj Singh Parmar}
	\author[1, $\S$]{Debasish Saha}
	\author[1,*]{Ranjini Bandyopadhyay}
	\date{\today}
	
	\footnotetext[2]{palak@rri.res.in}
	\footnotetext[3]{vaibhav@rri.res.in}
	\footnotetext[4]{Debasish.Saha@uni-duesseldorf.de}
	\footnotetext[1]{Corresponding Author: Ranjini Bandyopadhyay; Email: ranjini@rri.res.in}
	\maketitle
	\begin{abstract}
		Intricate fluid displacement patterns, arising from the unstable growth of interfacial perturbations, can be driven by fluid viscoelasticity and surface tension. A soft glassy suspension ages, $i.e.$ its mechanical moduli evolve with time, due to the spontaneous formation of suspension microstructures. The shear and time-dependent rheology of an aging suspension can be exploited to generate a wide variety of interfacial patterns during its displacement by a Newtonian fluid. Using video imaging, we report a rich array of interfacial pattern morphologies: dense viscous, dendritic, viscoelastic fracture, flower-shaped, jagged and stable, during the miscible and immiscible displacements of an aging colloidal clay suspension by Newtonian fluids injected into a radial quasi-two-dimensional geometry at different flow rates. We propose a new parameter, the areal ratio, which we define as the fully-developed pattern area normalized by the area of the smallest circle enclosing it. We show that the natural logarithms of the areal ratios uniquely identify the distinct pattern morphologies, such that each pattern can be segregated in a three-dimensional phase diagram spanned by the suspension aging time, the displacing fluid flow rate, and interfacial tension. Besides being of fundamental interest, our results are useful in predicting and controlling the growth of interfaces during fluid displacements. 
	\end{abstract}
	\definecolor{black}{rgb}{0.0, 0.0, 0.0}
	\definecolor{red(ryb)}{rgb}{1.0, 0.15, 0.07}
	\definecolor{darkred}{rgb}{0.55, 0.0, 0.0}
	\definecolor{blue(ryb)}{rgb}{0.01, 0.2, 1.0}
	\definecolor{darkcyan}{rgb}{0.0, 0.55, 0.55}
	\definecolor{navyblue}{rgb}{0.0, 0.0, 0.5}
	\definecolor{olivedrab(web)(olivedrab3)}{rgb}{0.42, 0.56, 0.14}
	\definecolor{darkraspberry}{rgb}{0.53, 0.15, 0.34}
	\definecolor{magenta}{rgb}{1.0, 0.0, 1.0}
	
	\newcommand{\blsquare}{\textcolor{black}{\small$\blacksquare$}}
	\newcommand{\hlsquare}{\textcolor{darkred}{\small$\square$}}
	\newcommand{\redtraingle}{\textcolor{magenta}{\small$\triangle$}}
	\newcommand{\oolive}{\textcolor{olivedrab(web)(olivedrab3)}{\large$\circ$}}
	\newcommand{\purpletraingle}{\textcolor{darkraspberry}{\small$\triangledown$}}
	
	\newcommand{\bltriangle}{\textcolor{black}{\small$\triangleup$}}
	\newcommand{\rcircle}{\textcolor{red(ryb)}{\large$\bullet$}}
	\newcommand{\rtraingle}{\textcolor{red(ryb)}{\small$\triangledown$}}
	\newcommand{\wine}{\textcolor{darkred}{\large$\bullet$}}
	\newcommand{\cyan}{\textcolor{darkcyan}{\large$\bullet$}}
	\newcommand{\owine}{\textcolor{darkred}{\large$\circ$}}
	\newcommand{\redcircle}{\textcolor{red}{\large$\circ$}}
	\newcommand{\ocyan}{\textcolor{darkcyan}{\large$\circ$}}
	\newcommand{\blue}{\textcolor{blue(ryb)}{\large$\bullet$}}
	\newcommand{\blbullet}{\textcolor{navyblue}{\large$\bullet$}}
	\newcommand{\olbullet}{\textcolor{olivedrab(web)(olivedrab3)}{\large$\bullet$}}
	\newcommand{\hollowblue}{\textcolor{blue(ryb)}{\large$\circ$}}
	\newcommand{\black}{\textcolor{black}{\large$\bullet$}}
	\newcommand{\hollowblack}{\textcolor{black}{\large$\circ$}}
	\newcommand{\bltria}{\textcolor{black}{\small$\triangle$}}
	\newcommand{\rtrai}{\textcolor{red(ryb)}{\large$\triangledown$}}
	\section{Introduction}
	
	The intrusion of one fluid into another can result in a wide range of interfacial patterns that depend on surface tension gradients~\cite{chen1989growth,bischofberger2015island}, viscoelasticities of the fluids~\cite{kondic1998non,park1994viscous,PALAK2021127405,lemaire1991viscous,ozturk2020flow,BALL2021104492,OSEIBONSU2016288},
	the driving pressures imposed during their injection~\cite{proud2005fractal}, etc. Unstable interfaces arise from the nonlinearities inherent in the system and have several implications in the technological realm, for example, in flows through porous media~\cite{Porous}, microfluidics~\cite{Q}, sugar refining~\cite{hill1952channeling}, enhanced oil recovery~\cite{orr1984use} and geophysics~\cite{perugini2005development,roy1996patterns}. A Hele-Shaw cell, a setup comprising two parallel plates typically separated by a gap of a few hundred micrometers, is often used to study confined flows~\cite{PINILLA2021e07614}. Such confined flows are governed by Darcy's law, with flow velocity depending upon the fluid pressure gradient, the viscosity of the fluid and the gap between the plates. Interfacial patterns resulting from the displacement of one Newtonian fluid by another were reported to be sensitive to the viscosity contrast between the fluid pair~\cite{bischofberger2014fingering}, the interfacial tension~\cite{suzuki2020experimental} and the pressure gradient~\cite{proud2005fractal}.
	\par
	However, instabilities that arise due to the displacement of a non-Newtonian fluid by a Newtonian fluid result in interfacial patterns that are quite distinct from those involving Newtonian-Newtonian displacements~\cite{park1994viscous,PALAK2021127405,lemaire1991viscous,ozturk2020flow,kondic1998non}. Numerical simulations revealed that the displacement of shear-thinning fluids by air resulted in the narrowing of fingers, multiple side branches and a suppression in finger-tip splitting~\cite{kondic1998non}. When foam, an yield stress fluid, was displaced by air at increasing shear rates, a transition from jagged to smooth interfacial patterns, believed to arise from the distortions and rearrangements of foam bubbles at high shear rates, was observed experimentally~\cite{park1994viscous}. Recent work on the displacement of dense granular cornstarch suspensions by glycerol-water mixtures reported a significant suppression of interfacial instabilities when either the viscosity ratio of the fluid pair or the concentration-dependent elasticity of the displaced viscoelastic suspension was increased~\cite{PALAK2021127405}. Apart from displacements of granular suspensions by Newtonian fluids, the withdrawal of polyethylene-silicone oil granular mixtures also resulted in a series of interfacial morphologies depending on the particle size, the particle aggregation structures, the hydrodynamic condition and the withdrawal rate of the mixture~\cite{LI20221598}. The radial displacement of Bentonite clay suspensions by water resulted in viscoelastic fracturing when the stored elastic energy of the displaced suspension exceeded the energy required for fracture onset~\cite{lemaire1991viscous}. Given the vital roles that clay suspensions play in the formation of river networks~\cite{roy1996patterns} and drilling applications~\cite{drilling}, a systematic study of their displacements is of paramount importance.
	\par
	Synthetic clay particles of  Laponite\textsuperscript{\textregistered} are disk-shaped with diameters 25-30 nm and thickness $\approx$ 1 nm. Dry Laponite powder exists in the form of one-dimensional stacks called tactoids with Na{$^+$} ions residing in the intergallery spaces of the clay platelets~\cite{van1977introduction}. In an aqueous suspension of pH < 11, the rims of the Laponite particles bear positive charges~\cite{pH}. The Laponite faces acquire negative charges due to the diffusion of Na{$^+$} ions into the aqueous medium~\cite{van1977introduction,lagaly2013handbook}. These Na{$^+$} counterions contribute to the formation of diffuse electric double layers around the Laponite particles. Osmotic pressure differences within the suspension cause tactoid swelling and the eventual exfoliation of Laponite platelets from the ends of the tactoids~\cite{bandyopadhyay2004evolution,misra}. This results in a time-dependent inter-particle electrostatic potential, such that the heterogeneously charged Laponite particles gradually self-assemble to form fragile microstructures composed of overlapping coins and house of cards (HoC) aggregates~\cite{delhorme2012monte}. The percolation of these microstructures throughout the sample is accompanied by continuous particle rearrangements within the fragile suspension structures. These processes contribute to a spontaneous temporal increase in sample rigidity in a phenomenon identified as physical aging~\cite{ali2016effect}.
	\begin{figure}[!b]
		\includegraphics[width= 6.0in]{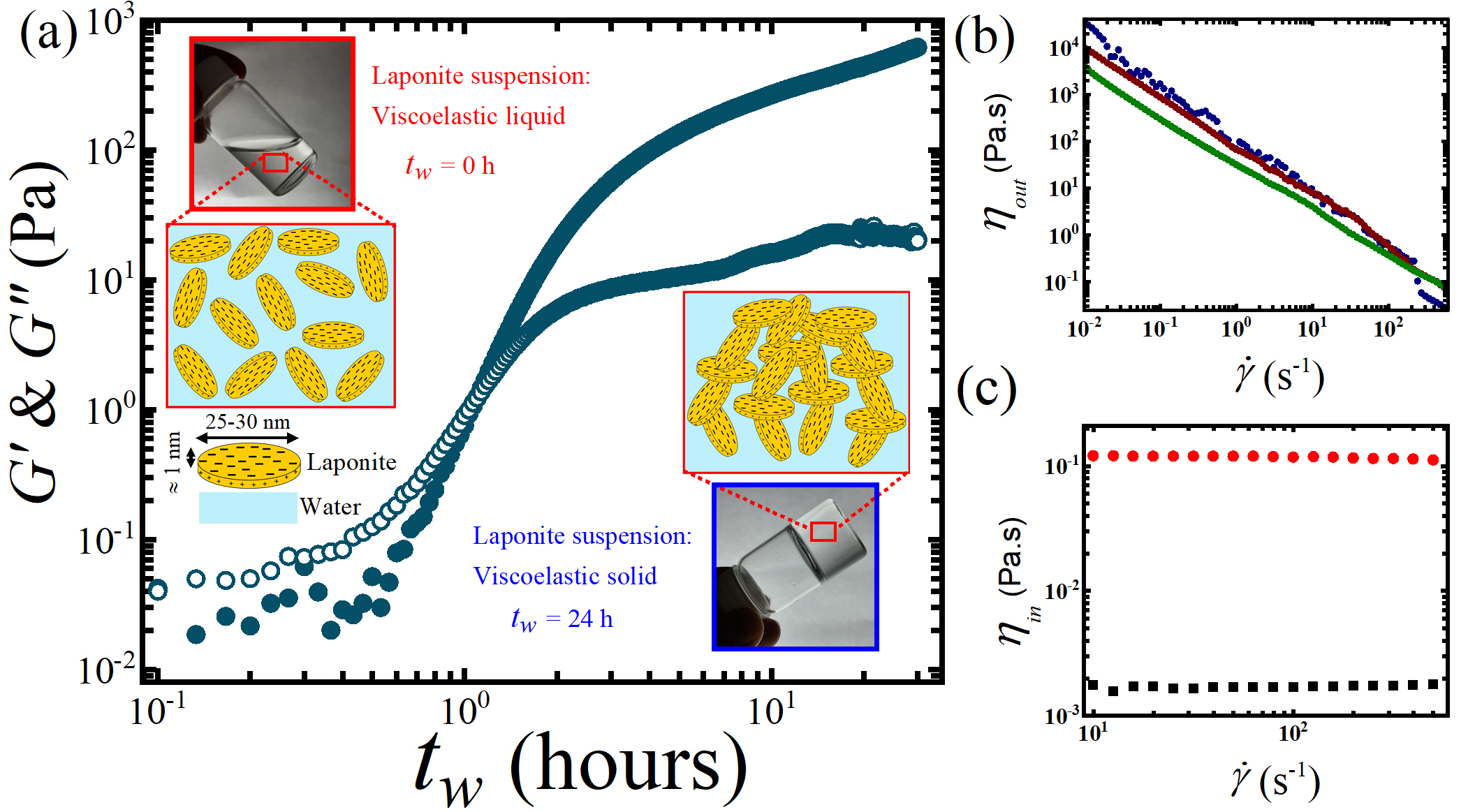}
		\centering
		\caption{\label{fig:rheo}\textbf{Rheological measurements.} \textbf{(a)} Evolution of elastic modulus $G^{\prime}$ (\cyan) and viscous modulus $G^{\prime \prime}$ (\ocyan) for a 3.25\% w/v Laponite suspension with increasing aging time $t_w$. The inset on the top-left shows a photograph of a freshly prepared aqueous clay suspension in a glass vial and a schematic illustration of well dispersed Laponite clay particles. The inset on the bottom-right shows a photograph of the same sample, acquired after $t_w = 24$ h, and a schematic representation of the house of cards (HoC) suspension microstructures formed by self-assembly of Laponite clay particles. \textbf{(b)} Viscosity $vs.$ shear rate ($\dot{\gamma}$) flow curves for 3.25\% w/v Laponite suspensions at aging times $t_w=$ 1 h (\olbullet), 5 h (\wine) and 24 h (\blbullet) show shear-thinning rheology. \textbf{(c)} Viscosity $vs.$ shear rate ($\dot{\gamma}$) curves for water (\blsquare) and mineral oil (\rcircle) display Newtonian flow.}
	\end{figure}  
	It was reported that the aging of aqueous Laponite suspensions can be controlled by changing clay concentration and suspension temperature, and by incorporating various additives in the dispersion medium~\cite{misra,Venketesh,Shahin2012,saha2015dynamic}. The inclusion of dissociating additives modifies interparticle electrostatic interactions, thereby altering the aging dynamics of clay suspension~\cite{saha2015dynamic}. Non-dissociating additives, in contrast, alter the structure of the dispersion medium. The resultant changes in intermolecular solvation forces have a non-trivial influence on sample aging~\cite{misra}.
	
	\par
	Rheological measurements such as in Fig.~\ref{fig:rheo}(a) can be employed to monitor the spontaneous evolution of the mechanical moduli of a viscoelastic clay suspension with aging time $t_w$; here $t_w$ is the time duration between sample loading and measurement. Additional details about rheological measurements are provided in Supplementary Material section ST1. We see from the photos in the insets of Fig.~\ref{fig:rheo}(a) that while a clay suspension at an early age shows liquid-like rheology (top-left inset), the same suspension, when aged considerably, behaves like a solid and supports its own weight against gravity (bottom-right inset). Schematic illustrations of the arrangement of Laponite particles in the aqueous suspension medium are also displayed. While the platelets are well-dispersed in the medium at low suspension ages (top-left inset), the clay particles self-assemble to form fragile percolating microstructures at higher ages (bottom-right inset). Investigations into the phase behavior of Laponite suspensions have reported phase separation below a concentration of 1 \% w/v~\cite{ruzicka2011fresh}, structural buildup and aging, presumably through Laponite house of cards (HoC) associations, between 1 \% w/v and 2 \% w/v~\cite{Jabbari,NICOLAI200151} and formation of a colloidal glassy phase above 2 \% w/v~\cite{Angelini2014}.
	
	We identify and analyze a wide range of interfacial patterns obtained by the displacement of an aging viscoelastic Laponite clay suspension by a Newtonian fluid in a radial Hele-Shaw cell. We study both miscible and immiscible displacements by using water and mineral oil as the displacing Newtonian fluids. As seen in Fig.~\ref{fig:rheo}(b), Laponite suspensions are strongly shear-thinning. In contrast, as shown in Fig.~\ref{fig:rheo}(c), water and mineral oil show Newtonian flow and their viscosities are independent of the imposed shear rates. We report three distinct interfacial pattern morphologies: dense viscous patterns (DVP), dendritic patterns (DP) and viscoelastic fractures (VEF) when an aqueous clay suspension of increasing $t_w$ and therefore having progressively larger elasticities, is displaced by water, a miscible solvent. In contrast, flower patterns (FP) transform to irregularly-shaped jagged patterns (JP) when an aqueous clay suspension of increasing $t_w$ is displaced by immiscible mineral oil. We note that the observed pattern morphologies are determined by the spontaneous aging and shear-thinning rheology of the displaced clay suspension, and the interfacial tension between the fluid pair. We quantify all the observed interfacial patterns by computing their average branch frequencies and average branch widths. Finally, we propose a new parameter, an areal ratio $A_p/A$, which we define as the area of the fully developed interfacial pattern, $A_p$, divided by the area of the smallest circle enclosing it, $A$. Estimation of ln($A_p/A$) allows us to segregate all the distinct morphologies in a three-dimensional nonequilibrium phase diagram spanned by the aging time ($t_w$) of the displaced clay suspension, the flow rate ($q$) of the displacing Newtonian fluid and the interfacial tension ($\sigma$).

	\section{\label{em}Materials and Methods}

	\begin{figure}[!b]
		\includegraphics[width= 6.0in]{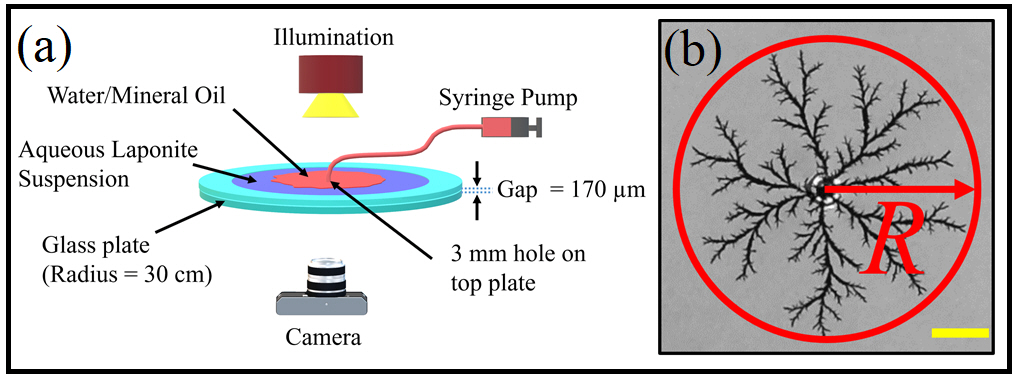}
		\centering
		\caption{\label{fig:hssem}\textbf{Experimental setup.} \textbf{(a)} Schematic diagram showing the side view of a radial Hele-Shaw cell. \textbf{(b)} A representative grayscale displacement pattern of water (black) invading an aging aqueous clay suspension (gray). $R$ is the longest finger length of the pattern such that a circle of radius $R$ encloses the entire pattern. The scale bar is 5 cm.}
	\end{figure}
	Dried Laponite XLG powder (BYK additives Inc.) was vigorously stirred in Milli-Q water (Millipore Corp., resistivity 18.2 MΩ.cm) to prepare a 3.25\% w/v Laponite suspension. The freshly prepared Laponite clay suspension was first loaded in a radial Hele-Shaw cell, comprising two circular glass plates each of radius 30~cm separated by a gap of $170~\mu m$, with a syringe pump (NE-8000, New Era Pump Systems, USA) through a 3 mm central hole drilled on the top plate (Fig.~\ref{fig:hssem}(a)). The aging time $t_w$ = 0 corresponds to the time when the loading of the clay suspension was completed. Milli-Q water and mineral oil (Acros Organics) were used as the miscible and immiscible displacing fluids respectively in two different sets of experiments. These Newtonian fluids were injected with flow rate $q$ through the same central hole of the Hele-Shaw cell after allowing the clay suspension to age to a pre-determined aging time $t_w$. The displacement an aging clay suspension by a Newtonian fluid resulted in interfacial patterns whose evolutions were recorded. The stack of images obtained from the videos was converted to grayscale format (Fig.~\ref{fig:hssem}(b)) and analyzed using the MATLAB@2018 image processing toolbox. A stress-controlled rheometer (Anton Paar, MCR 501)  was used to perform rheological measurements in a cone and plate geometry. All the experiments were performed at room temperature (25$^{\circ}$C). Additional experimental details are provided in Supplementary Material section ST1. Raw images of representative interfacial patterns are shown in Supplementary Fig.~S1. The procedure for binarization of raw images is provided in Supplementary Fig.~S2.
	
	\section{\label{r&d}Results and Discussion}
	\subsection{\label{r&d:eowt}Aging time of the displaced suspension determines interfacial pattern morphology}
	\begin{figure}[!t]
		\includegraphics[width=6.0in]{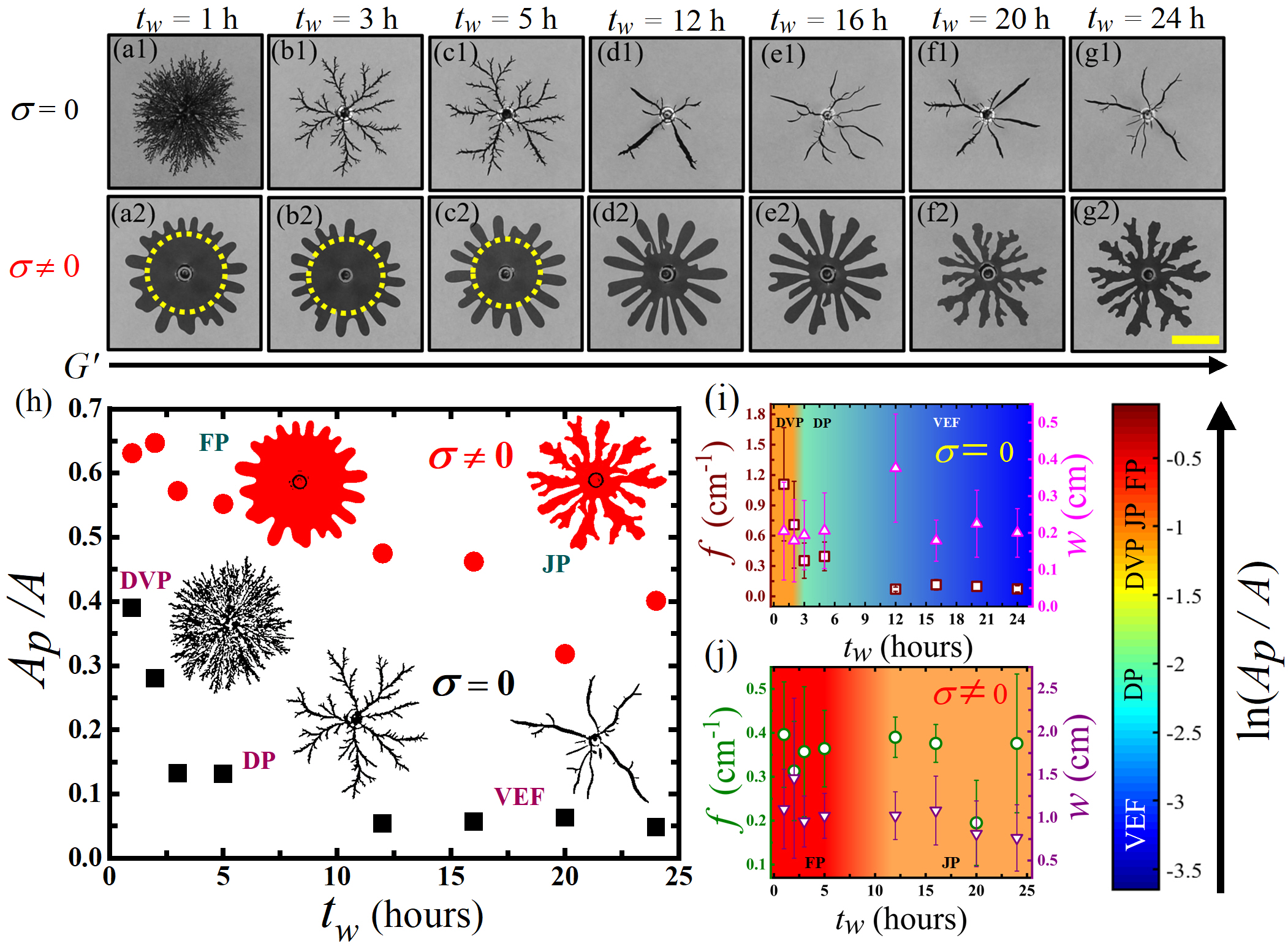}
		\centering
		\caption{\label{agingtime}\textbf{Pattern morphologies and their characterization at different ages, $\boldsymbol{t_w}$, of the displaced suspension.} Miscible (interfacial tension $\sigma= 0$) displacement patterns in grayscale obtained by the displacement of a 3.25\% w/v aqueous clay suspension by water at constant flow rate $q = 5$ ml/min and for different suspension aging times $t_w$: \textbf{(a1)} dense viscous patterns (DVP), \textbf{(b1-c1)} dendritic patterns (DP) and \textbf{(d1-g1)} viscoelastic fractures (VEF). Immiscible ($\sigma \neq 0$ ) displacement patterns obtained by the displacement of an aging clay suspension at the same $t_w$ values and $q = 5$ ml/min: \textbf{(a2-c2)} flower patterns (FP). Yellow dotted circles illustrate the central dark stable region. \textbf{(d2-g2)} jagged patterns (JP). All the patterns correspond to a fixed longest finger length $R$ = 7.01 $\pm$ 0.01 cm. The scale bar is 4 cm. \textbf{(h)} The areal ratio $A_p/A$ of the patterns vs. $t_w$ for miscible (\blsquare) and immiscible (\rcircle) displacements. Here, $A_p$ is the area of the fully developed pattern and $A = \pi R^2$ is the area of the smallest circle enclosing the entire pattern. \textbf{(i)} Average branch frequency ($f$, \hlsquare ) and average width ($w$, \redtraingle) of the fingers as a function of $t_w$ for miscible ($\sigma = 0$) displacement patterns. \textbf{(j)} $f$ (\oolive) and $w$ (\purpletraingle) vs. $t_w$ for immiscible ($\sigma \neq 0$) displacement patterns. Error bars in $f$ and $w$ represent the standard deviations in measurements of multiple branches. The vertical color bar on the right maps the distinct morphological patterns to the natural logarithms of their areal ratios: ln($A_p/A$).}
	\end{figure}
	\par
	Figures~\ref{agingtime}(a1-g1, a2-g2) show representative interfacial patterns formed during the radial displacement of an aqueous clay suspension of increasing ages, $t_w$, by miscible and immiscible Newtonian displacing fluids at a constant flow rate $q = 5$ ml/min. All the interfacial patterns correspond to a fixed longest finger length $R$ = 7.01 $\pm$ 0.01 cm ($R$ is defined in Fig.~\ref{fig:hssem}(b)). We note three distinct morphologies for miscible displacements by water (interfacial tension $\sigma$ = 0) as the elasticity of the displaced clay suspension increases due to aging (Figs.~\ref{agingtime}(a1-g1)). For a clay suspension of low age ($t_w$ = 1 h), we observe the frequent merging of neighboring fingers during pattern evolution, resulting in a pattern without well-defined boundaries (Fig.~\ref{agingtime}(a1)). We designate this pattern, characterized by narrow fingers and side branches, as a dense viscous pattern (DVP, Supplementary Movie 1). At intermediate $t_w$, we observe dendritic patterns (DP, shown in Figs.~\ref{agingtime}(b1,c1), Supplementary Movie 2) with considerable side-branching and well-defined interfacial boundaries. For highly aged suspensions characterized by large $t_w$ values, viscoelastic fractures (VEF, Supplementary Movie 3) emerge (Figs.~\ref{agingtime}(d1-g1)) and propagate rapidly. The VEF patterns display sharp finger-tips and perpendicular offshoots from the primary cracks, and are morphologically very dissimilar to DVP and DP.
	\par
	The observed side-branching in DVP and DP occurs due to flow-induced anisotropies~\cite{Anisotropy} arising from shear-thinning of the increasingly elastic displaced clay suspension. Interestingly, nonlinear simulations~\cite{kondic1998non} revealed that a competition between enhanced finger-tip velocity due to shear-thinning of a displaced viscoelastic fluid and simultaneous finger-tip broadening due to imposed driving pressures can result in the formation of side branches. Here, we estimate the finger-tip velocity of each pattern by recording the evolution of the longest finger of length $R$ with time. With the transition of the patterns from DVP to VEF, the time-averaged values of finger-tip velocities ($\bar{U}$, Supplementary Fig.~S3(a)) increase. These velocity values are used to calculate the shear rates ($\dot\gamma = 2\bar{U}/b$ where $b$ is the Hele-Shaw cell gap~\cite{nagatsu_matsuda_kato_tada_2007}) imposed by the displacing fluid. The estimated values of shear rates at the finger-tips ($36.47~\mathrm{s^{-1}}\leq\dot{\gamma}\leq552.94~\mathrm{s^{-1}}$) increase due to enhanced shear-thinning of the displaced clay suspension as its age, $t_w$, increases. This leads to the increased shedding of side branches. However, for clay suspensions having large $t_w$, elastic effects dominate over shear-thinning, resulting in the emergence of VEF patterns. Viscoelastic fractures were reported in earlier experiments involving miscible displacements of elastic Bentonite clay suspensions~\cite{lemaire1991viscous}. Such patterns were also seen when elastic agar gel suspensions~\cite{hirata1998fracturing} and shear-jammed cornstarch suspensions~\cite{ozturk2020flow} were displaced by immiscible air under large driving pressures. In our experiments with the highly aged clay suspension, the system-wide percolation of suspension microstructures causes the buildup of elastic stresses. Local yielding events in the jammed clay suspension during its displacement by water lead to the rapid release of elastic stresses and the observed emergence of viscoelastic fractures (VEF).
	
	\par
	Since the formation of sharp tips costs more energy when the interfacial tension is non-zero, immiscible displacement patterns preferentially preserve smooth and rounded finger-tips. The stabilizing effect of interfacial tension results in considerably lower finger-tip velocities $\bar{U}$ (Supplementary Fig.~S3(b)) and shear rates ($12.94~\mathrm{s^{-1}}\leq\dot{\gamma}\leq16.47~\mathrm{s^{-1}}$) than in miscible displacement experiments (Supplementary Fig.~S3(a)). Furthermore, for immiscible displacements, $\bar{U}$ does not show a significant dependence on the age of the displaced clay suspension. We note that immiscible displacement patterns (Figs.~\ref{agingtime}(a2-g2)) have very distinct morphologies when compared to their miscible counterparts at identical suspension ages and displacing fluid flow rates. Flower patterns (FP, Supplementary Movie 4) with round finger-tips and a dark central region (yellow dotted circles in Figs.~\ref{agingtime}(a2-c2)) are observed when suspensions of low to intermediate ages are displaced by mineral oil. This dark central region results from a delay in onset of instability at non-zero interfacial tension and resembles the observations in previous experiments involving Newtonian fluid pairs~\cite{paterson1981radial,bischofberger2015island}. Displacement of a highly aged clay suspension having large $t_w$ requires disruption of the fragile suspension microstructures by a process that costs substantial energy and leads to a transition to more ramified jagged patterns (JP, shown in Figs.~\ref{agingtime}(d2-g2), Supplementary Movie 5). It is to be noted that the observed jagged patterns closely resemble those seen during the immiscible displacement of shaving foam, a non-zero yield stress material, by air~\cite{park1994viscous}. 
	
	\par
	We quantify the morphologies of the interfacial patterns (Figs.~\ref{agingtime}(a1-g1, a2-g2)) by estimating their areal ratios ($A_p/A$), average branch frequencies ($f$) and average branch widths ($w$). Figure~\ref{agingtime}(h) shows the areal ratios, defined as the area of the fully developed pattern ($A_p$) normalized by the area of the smallest circle enclosing it ($A = \pi R^2$, Fig.~\ref{fig:hssem}(b)), as a function of aging time $t_w$. As a result of the wide branching morphologies of the immiscible patterns (Figs.~\ref{agingtime}(a2-g2)), we note that their areal ratios $A_p/A$ are always greater than for the miscible ones at the same suspension age. The measurements of average branch frequencies are performed by analyzing a narrow annular region close to the outermost boundary of the fully developed interfacial patterns, as displayed in Supplementary Fig.~S4. The average branch frequency is defined as $f$ = $<N/2 \pi r>_r$, where $N$ is the number of branches intersecting a circle of radius $r$ and $<>_r$ denotes the average over circles of different radii in the annulus of interest. As seen in Fig.~\ref{agingtime}(i), viscoelastic fractures (VEF) display smaller branch frequencies when compared to dendritic patterns (DP) and dense viscous patterns (DVP). The average finger width is also estimated from measurements within the same annulus. We note that the thick, uniform fingers that form in immiscible displacement experiments (Fig.~\ref{agingtime}(j)) point to the critical role that surface tension plays in stabilizing flow instabilities. We conclude that the interfacial tension between the fluid pair and the aging time of the displaced suspension are reliable control parameters in determining elasticity-induced changes in the morphological features of miscible and immiscible displacement patterns.
	
	\begin{figure}[!t]
		\includegraphics[width=6.0in]{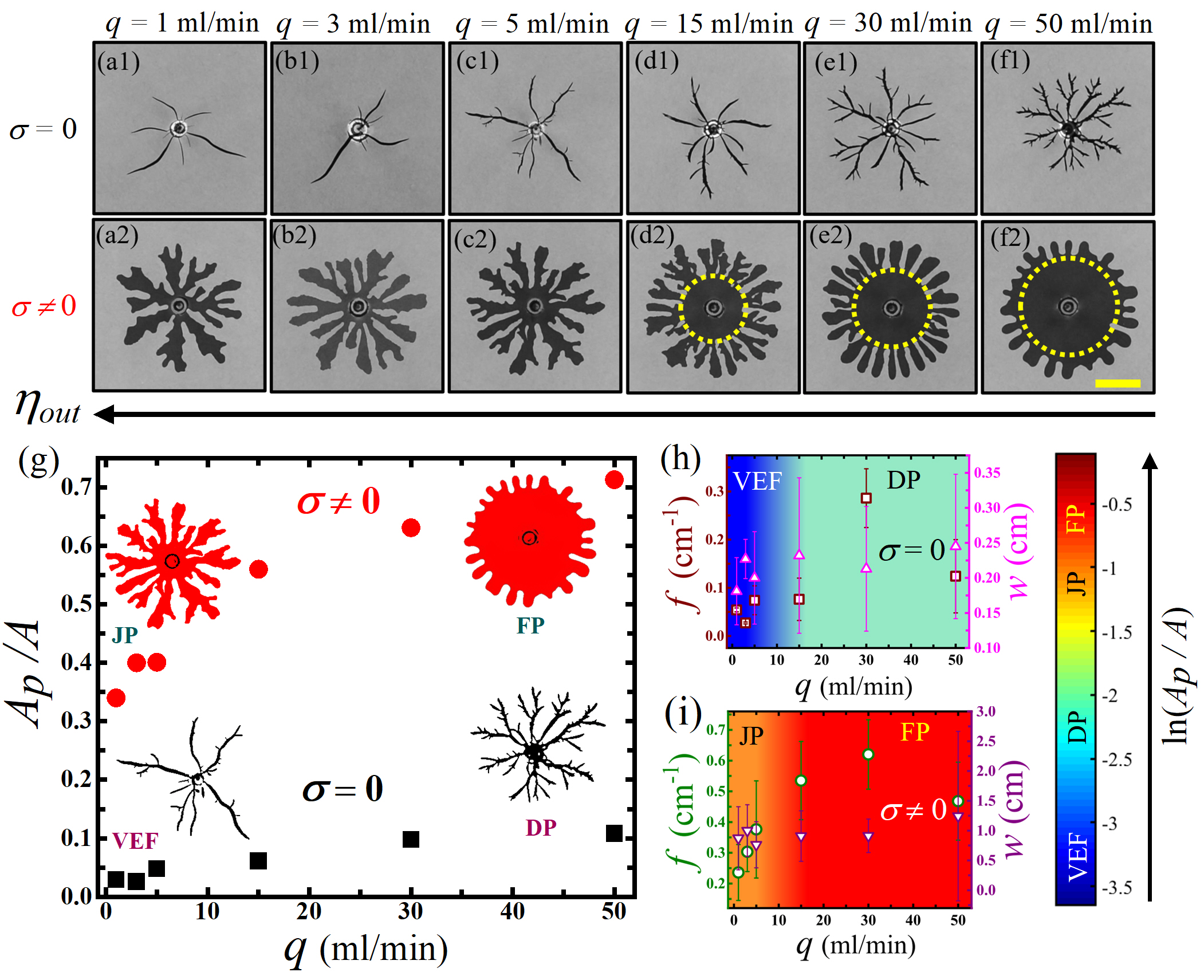}
		\centering
		\caption{\label{flowrate}\textbf{Pattern morphologies and their characterization at different displacing fluid flow rates, $\boldsymbol{q}$.} Miscible displacement patterns in grayscale, obtained by the displacement of a 3.25\% w/v aqueous clay suspension of $t_w$ = 24 h by water at various flow rates $q$: \textbf{(a1-c1)} viscoelastic fractures (VEF) and \textbf{(d1-f1)} dendritic patterns (DP). Immiscible displacement patterns at the same flow rates as in (a1-f1): \textbf{(a2-c2)} jagged patterns (JP) and \textbf{(d2-f2)} flower patterns (FP). All the patterns have a fixed longest finger length $R$ = 7.01 $\pm$ 0.01 cm. The scale bar is 4 cm. \textbf{(g)} $A_p/A$ as a function of flow rates for miscible (\blsquare) and immiscible (\rcircle) displacements. \textbf{(h)} Average branch frequency ($f$, \hlsquare ) and average width ($w$, \redtraingle) of the fingers as a function of $q$ for miscible ($\sigma = 0$) displacement patterns. \textbf{(i)} $f$ (\oolive) and $w$ (\purpletraingle) vs. $q$ for immiscible ($\sigma \neq 0$) displacement patterns. The distinct morphological regimes are
			mapped on a color bar using the natural logarithms of their areal ratios: ln($A_p/A$).}
	\end{figure}
	\subsection{\label{r&d:eoef}Pattern morphology can be tuned by controlling the shear-thinning rheology of the displaced suspension}
	
	The viscous modulus of a 3.25\% w/v aqueous clay suspension plateaus at $t_w$ $\approx$ 24 h, while its elastic modulus continues to evolve (Fig.~\ref{fig:rheo}(a)). We next explore the radial displacement of a sufficiently elastic suspension of age $t_w=24$ h in the Hele-Shaw cell due to the injection of Newtonian displacing fluids at different flow rates $q$. Viscoelastic fractures (VEF) with perpendicular offshoots are observed when water displaces the elastic clay suspension at low flow rates (Figs.~\ref{flowrate}(a1-c1)). The low driving pressures cause minimal shear-thinning of the displaced clay suspension such that fractures propagate rapidly due to the release of elastic stresses. On increasing the flow rate, VEF are replaced by more ramified dendritic patterns (DP) with pointed tips that display multiple splitting events during their growth (Figs.~\ref{flowrate}(d1-f1)). The average finger-tip velocities ($\bar{U}$) increase with increasing flow rates of the displacing fluid (Supplementary Fig.~S5(a)). The shear rates ($158.82~\mathrm{s^{-1}}\leq\dot{\gamma}\leq1907.05~\mathrm{s^{-1}}$) imposed at the finger-tip by the displacing fluid, estimated using $2\bar{U}/b$~\cite{nagatsu_matsuda_kato_tada_2007} as discussed earlier, therefore also increase with increasing flow rates, resulting in considerable shear-thinning of the displaced clay suspension. At high flow rates of the displacing fluid, the fragile microstructures formed by the self-assembling Laponite particles rupture irreversibly. This reduces suspension elasticity and results in the observed change in interfacial pattern morphology from VEF to DP. 
	
	\par
	The immiscible displacement of a clay suspension with mineral oil leads to a transition from jagged patterns (JP) to flower patterns (FP) as the flow rate of the displacing mineral oil is increased (Figs.~\ref{flowrate}(a2-f2)). The finger-tip velocities ($\bar{U}$) and the corresponding shear rates ($4.70~\mathrm{s^{-1}}\leq\dot{\gamma}\leq82.35~\mathrm{s^{-1}}$) imposed by the mineral oil increase with increasing flow rate but are always less than those noted in miscible displacement experiments (Supplementary Figs.~S5(a,b)). In order to understand the observed flow stabilization against viscous fingering instabilities at increasing flow rates, we estimate the viscosity ratios of the fluid pair $\eta_{in}/\eta_{out}$, where $\eta_{in}$ is the viscosity of the displacing Newtonian fluid and $\eta_{out}$ is the shear-dependent viscosity of the displaced clay suspension at different flow rates $q$. The shear-thinning of the displaced clay suspension at high shear rates leads to an increase in the viscosity ratio of the fluid pair (Supplementary Material section ST2). This reduces the destabilizing action of viscous forces and results in increasingly stable interfaces at high injection flow rates. Such suppression of viscous instabilities has been noted in previous work involving Newtonian-Newtonian~\cite{bischofberger2014fingering,lajeunesse19973d} and Newtonian-non-Newtonian displacements~\cite{PALAK2021127405}. Since mineral oil is more viscous than water (Figs.~\ref{fig:rheo}(c)), the increase in viscosity ratio with increasing $q$ is more pronounced in immiscible displacements when compared to the miscible cases. This causes the observed flow stabilization and the transition from JP to FP when $q$ is increased (Figs.~\ref{flowrate}(a2-f2)). At the highest viscosity ratio (Supplementary Fig.~S5(c)), we observe a stable pattern (SP, Supplementary Fig.~S6) with a circularly growing interface at all times during the immiscible displacement of a clay suspension of low age ($t_w$ = 1 h) at a high flow rate ($q$ = 50 ml/min). We conclude from the data presented in Fig.~\ref{flowrate} that shear-thinning effects dominate as the injection flow rate is increased, and the flow stabilizes due to two factors, $viz.$, an increase in the viscosity ratio of the fluid pair and the presence of a non-zero interfacial tension. Pattern morphologies at a radial quasi-two-dimensional interface can therefore also be effectively tuned by controlling the flow rate of the displacing fluid.
	\begin{figure}[!t]
		\includegraphics[width=2.5in]{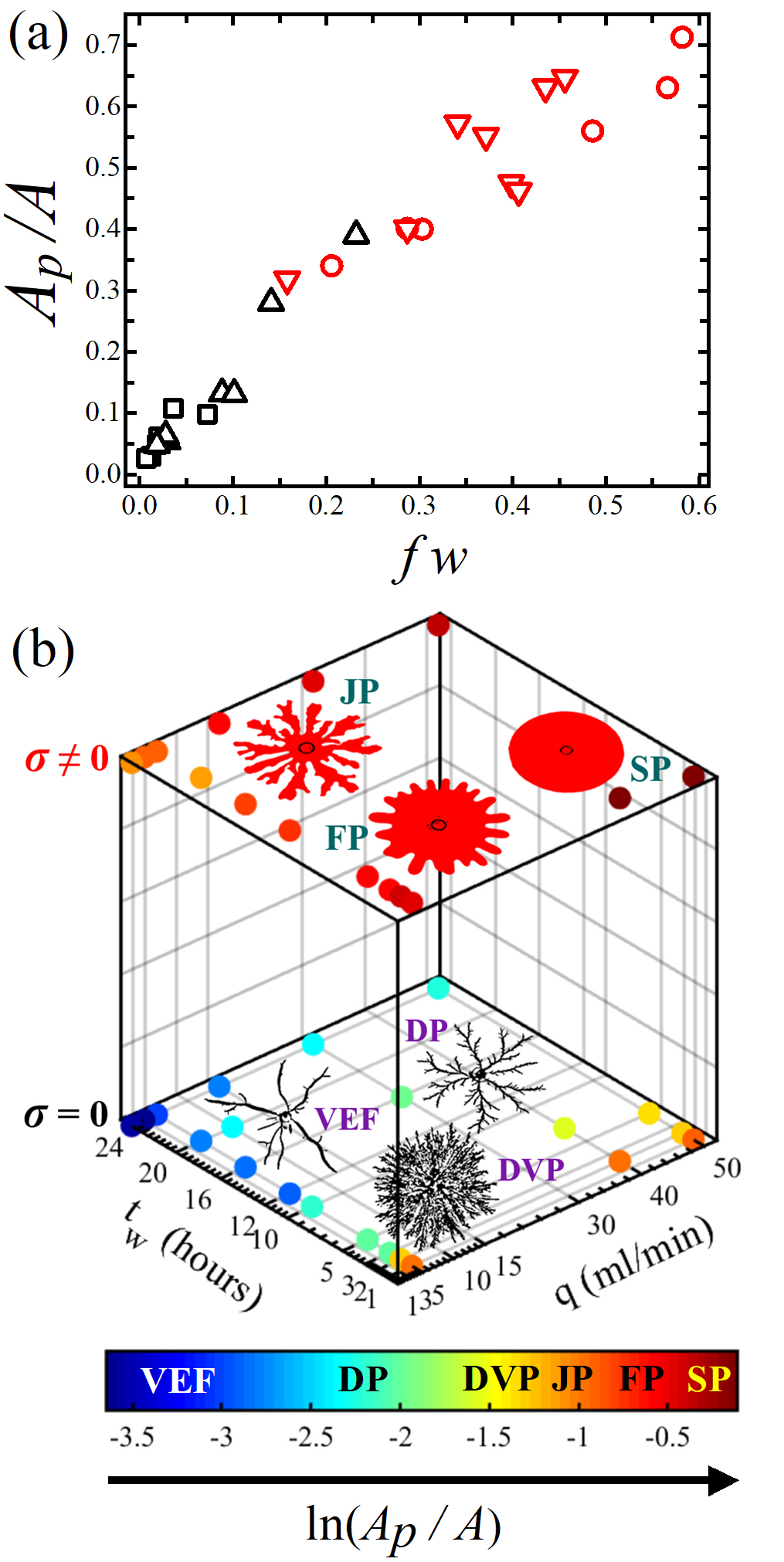}
		\centering
		\caption{\label{phase}\textbf{Segregation of pattern morphologies.} \textbf{(a)} $A_p/A$ vs.~$fw$ for miscible ($\triangle$ for variable $t_w$, $\square$ for variable $q$ ) and immiscible (\rtrai~for variable $t_w$, \redcircle~for variable $q$) displacements. \textbf{(b)} Phase diagram relating the suspension aging time $t_w$, the displacing fluid flow rate $q$,  and the interfacial tension $\sigma$: bottom plane shows miscible ($\sigma = 0$) displacements with three different pattern morphologies: dense viscous patterns (DVP), dendritic patterns (DP) and viscoelastic fractures (VEF). Top plane shows immiscible ($\sigma \neq 0$) displacement patterns with three different morphologies: flower patterns (FP), jagged patterns (JP) and stable patterns (SP). The colored solid circles represent all the distinct interfacial patterns observed experimentally. The colors of these circles, defined in the colorbar at the bottom, have one-to-one mapping with the natural logarithms of the pattern areal ratios: ln($A_p/A$).}
	\end{figure}
	\par
	We next quantify the morphologies of the observed interfacial patterns in terms of their areal ratios ($A_p/A$), average branch frequencies ($f$) and average branch widths ($w$). As in our earlier experiments in which we displaced a clay suspension at pre-determined aging times, $t_w$ (Fig.~\ref{agingtime}(h)), $A_p/A$ is larger for immiscible displacement patterns when compared to the miscible ones at a fixed flow rate of the displacing fluid (Fig.~\ref{flowrate}(g)). Dendritic patterns (DP) obtained at high $q$ during miscible displacements display larger branch frequencies when compared to viscoelastic fractures (VEF). The presence of non-zero interfacial tension results in larger average branch widths $w$ in the immiscible displacement patterns (Fig.~\ref{flowrate}(i)) when compared to their miscible counterparts (Fig.~\ref{flowrate}(h)). The interfacial patterns obtained at different ages $t_w$ of the displaced clay suspensions (displayed in Figs.~\ref{agingtime}(a1-g1,a2-g2)) can therefore be recovered by keeping $t_w$ fixed while appropriately changing the flow rates $q$ of the displacing fluids (Figs.~\ref{flowrate}(a1-f2,a2-f2)). Interestingly therefore, displacing a clay suspension of increasing age at a fixed flow rate generates the same sequence of patterns as seen during the displacement of the suspension at a fixed age while lowering the flow rates of the displacing fluid.
	
	\par
	Regardless of whether the interfacial patterns were generated while changing the flow rate of the displacing fluid or the aging time of the displaced clay suspension, we see from Fig.~\ref{phase}(a) that their areal ratios ($A_p/A$) are linearly correlated with the products of their average branch frequencies ($f$) and widths ($w$). As noted earlier, we observe a stable pattern (SP) during the immiscible displacement of a clay suspension of low age ($t_w$ = 1 h) at a high flow rate ($q$ = 50 ml/min, Supplementary Figs.~S5(c), S6). Since it is not feasible to estimate the branch frequency and finger width of a stable pattern, we compute the natural logarithms of the areal ratios, ln($A_p/A$), of all the patterns and segregate them in a three-dimensional nonequilibrium phase diagram (Fig.~\ref{phase}(b)) spanned by the aging time ($t_w$) of the displaced clay suspension, the flow rate ($q$) of the displacing fluid and the interfacial tension ($\sigma$). All the interfacial patterns obtained in our experiments are represented on the phase diagram in Fig.~\ref{phase}(b) by symbols of distinct colors that are mapped to their respective ln($A_p/A$) values, as defined in the color bar. We therefore identify the natural logarithm of areal ratio ln($A_p/A$), as a single parameter that can be employed to effectively distinguish the wide range of quasi-two-dimensional interfacial patterns formed in our Newtonian-non-Newtonian displacement experiments.

	\section{\label{se:sac} Conclusions}
	
	Since a colloidal clay suspension is viscoelastic and shows spontaneous aging, its time-dependent elastic modulus and shear-thinning rheology should lead to the generation of a wide range of patterns when it is displaced at different flow rates by Newtonian fluids of contrasting miscibilities. In this work, we systematically study the selection of interfacial patterns by radially displacing an aging soft glassy aqueous clay suspension with miscible and immiscible Newtonian fluids in a confined geometry. We observe and analyze the wide variety of interfacial morphologies that are generated in these controlled experiments. We note that the observed patterns strongly depend on the elasticity of the displaced suspension, parameterized in terms of the suspension aging time, the injection flow rate of the displaced Newtonian fluid, and the miscibility of the fluid pair. For both miscible and immiscible displacements, we understand the transitions between the observed interfacial patterns by considering the competition between interfacial, elastic and viscous forces in the aging and shear-thinning clay suspension. We note that instabilities driven by viscous forces result in interfacial patterns characterized by tip-splitting, side branches, and thick protrusions resembling flower petals. In contrast, if elastic forces dominate due to the spontaneous formation of suspension microstructures~\cite{ali2016effect,delhorme2012monte,bandyopadhyay2004evolution}, we observe viscoelastic fracturing or jagged interfaces. In both scenarios, the presence of a non-zero interfacial tension leads to enhanced flow stabilization. Our work supplements the existing experimental literature on clay suspension displacements~\cite{lemaire1991viscous,van1986fractal,VanDamme1990} by identifying the suspension aging time $t_w$, which is closely correlated with suspension elasticity, as a new control parameter that reliably determines interfacial pattern morphologies. Besides extending our fundamental knowledge of viscoelastic flows in confined environments, the present work has important implications in understanding geophysical phenomena such as river delta formation~\cite{roy1996patterns}, and in controlling the displacements of viscoelastic materials such as mud and cement slurries~\cite{Bittleston2002}.
	
	\par
	The control of instabilities at the interface between a pair of fluids is crucial in material transport. Interfacial instabilities are essential in designing and optimizing several processes such as filtration and flow in porous media~\cite{Porous}. However, they are undesirable in certain scenarios, such as in the formation of dendrites at the anode-electrolyte interface in rechargeable lithium batteries~\cite{battery} and the emergence of viscous fingers at the water-oil interface in enhanced oil recovery~\cite{gorell1983theory}. Therefore, systematic suppression or enhancement of the interfacial instabilities is important, and different strategies to control these instabilities were reported earlier\cite{Gao2019, Li, Zheng,PALAK2021127405,borgman2019immiscible}. To the best of our knowledge, the research presented here is the first report of the sensitive dependence of the interfacial pattern morphology on the age of the displaced suspension. Our demonstration that the natural logarithms of the areal ratios can effectively distinguish pattern morphologies on a three-dimensional phase diagram can have far-reaching ramifications in controlling and predicting pattern morphologies between a fluid pair with well-characterized physicochemical properties. 
	\par
	Interparticle electrostatic interactions, and therefore the aging dynamics of a charged colloidal clay suspension, can be altered by varying clay concentration, temperature, and by incorporating dissociating additives such as salts and acids~\cite{saha2015dynamic,Venketesh,Shahin2012}. It would be of interest to systematically study the roles of diverse physicochemical conditions in the formation of interfacial patterns during the displacement of a clay suspension by miscible and immiscible Newtonian fluids. A possible extension of this work could involve the addition of non-dissociating molecules (kosmotropes such as glucose that enhance water structure~\cite{glucose}, or chaotropes such as N,N-Dimethylformamide that disrupt hydrogen bonding in water~\cite{dmf}) to the aqueous medium during preparation of the clay suspension~\cite{misra}. The presence of additives is expected to alter the aging dynamics and rheology of the suspension, which would strongly influence the interfacial pattern morphologies generated when the latter is displaced. While this paper focuses only on the distinct morphological features of fully developed interfacial patterns, we note that the time-evolution of the interface is quite different in each case. A complete understanding of the patterns that form at the interface between an aging clay suspension and a Newtonian fluid therefore requires high-speed video imaging of pattern onset and growth. Such work is ongoing in our laboratory.
	
	
	\section*{Data Availability}
	Source data are available for this paper from the corresponding author upon reasonable request.
	
	\section*{Acknowledgments}	 
	We thank Sreeram K. Kalpathy for useful discussions. Funding: We thank Raman Research Institute (RRI, India) for funding our research and Department of Science and Technology Science and Education Research Board (DST SERB, India) grant EMR/2016/006757 for partial support.
	
	\bibliographystyle{elsarticle-num}
	\bibliography{poly}

\begin{thebibliography}{10}
\expandafter\ifx\csname url\endcsname\relax
  \def\url#1{\texttt{#1}}\fi
\expandafter\ifx\csname urlprefix\endcsname\relax\def\urlprefix{URL }\fi
\expandafter\ifx\csname href\endcsname\relax
  \def\href#1#2{#2} \def\path#1{#1}\fi

\bibitem{chen1989growth}
J.-D. Chen, Growth of radial viscous fingers in a {Hele-Shaw} cell, J. Fluid
  Mech. 201 (1989) 223–242.
\newblock \href {https://doi.org/10.1017/S0022112089000911}
  {\path{doi:10.1017/S0022112089000911}}.

\bibitem{bischofberger2015island}
I.~Bischofberger, R.~Ramachandran, S.~R. Nagel, An island of stability in a sea
  of fingers: emergent global features of the viscous-flow instability, Soft
  Matter 11 (2015) 7428--7432.
\newblock \href {https://doi.org/10.1039/C5SM00943J}
  {\path{doi:10.1039/C5SM00943J}}.

\bibitem{kondic1998non}
L.~Kondic, M.~J. Shelley, P.~Palffy-Muhoray, {Non-Newtonian Hele-Shaw Flow and
  the Saffman-Taylor Instability}, Phys. Rev. Lett. 80 (1998) 1433--1436.
\newblock \href {https://doi.org/10.1103/PhysRevLett.80.1433}
  {\path{doi:10.1103/PhysRevLett.80.1433}}.

\bibitem{park1994viscous}
S.~S. Park, D.~J. Durian, Viscous and elastic fingering instabilities in foam,
  Phys. Rev. Lett. 72 (1994) 3347--3350.
\newblock \href {https://doi.org/10.1103/PhysRevLett.72.3347}
  {\path{doi:10.1103/PhysRevLett.72.3347}}.

\bibitem{PALAK2021127405}
Palak, R.~Sathyanath, S.~K. Kalpathy, R.~Bandyopadhyay, Emergent patterns and
  stable interfaces during radial displacement of a viscoelastic fluid,
  Colloids Surf. A Physicochem. Eng. Asp. 629 (2021) 127405.
\newblock \href
  {https://doi.org/https://doi.org/10.1016/j.colsurfa.2021.127405}
  {\path{doi:https://doi.org/10.1016/j.colsurfa.2021.127405}}.

\bibitem{lemaire1991viscous}
E.~Lemaire, P.~Levitz, G.~Daccord, H.~Van~Damme, From viscous fingering to
  viscoelastic fracturing in colloidal fluids, Phys. Rev. Lett. 67 (1991)
  2009--2012.
\newblock \href {https://doi.org/10.1103/PhysRevLett.67.2009}
  {\path{doi:10.1103/PhysRevLett.67.2009}}.

\bibitem{ozturk2020flow}
D.~Ozturk, M.~L. Morgan, B.~Sandnes, Flow-to-fracture transition and pattern
  formation in a discontinuous shear thickening fluid, Commun. Phys. 3~(1)
  (2020) 1--9.
\newblock \href {https://doi.org/10.1038/s42005-020-0382-7}
  {\path{doi:10.1038/s42005-020-0382-7}}.

\bibitem{BALL2021104492}
T.~V. Ball, N.~J. Balmforth, A.~P. Dufresne, Viscoplastic fingers and fractures
  in a {Hele-Shaw} cell, J. Non-Newton. Fluid Mech. 289 (2021) 104492.
\newblock \href {https://doi.org/https://doi.org/10.1016/j.jnnfm.2021.104492}
  {\path{doi:https://doi.org/10.1016/j.jnnfm.2021.104492}}.

\bibitem{OSEIBONSU2016288}
K.~Osei-Bonsu, N.~Shokri, P.~Grassia, Fundamental investigation of foam flow in
  a liquid-filled {Hele-Shaw} cell, J. Colloid Interface Sci. 462 (2016)
  288--296.
\newblock \href {https://doi.org/https://doi.org/10.1016/j.jcis.2015.10.017}
  {\path{doi:https://doi.org/10.1016/j.jcis.2015.10.017}}.

\bibitem{proud2005fractal}
O.~Praud, H.~L. Swinney, Fractal dimension and unscreened angles measured for
  radial viscous fingering, Phys. Rev. E 72 (2005) 011406.
\newblock \href {https://doi.org/10.1103/PhysRevE.72.011406}
  {\path{doi:10.1103/PhysRevE.72.011406}}.

\bibitem{Porous}
L.~Paterson, {Diffusion-Limited Aggregation and Two-Fluid Displacements in
  Porous Media}, Phys. Rev. Lett. 52 (1984) 1621--1624.
\newblock \href {https://doi.org/10.1103/PhysRevLett.52.1621}
  {\path{doi:10.1103/PhysRevLett.52.1621}}.

\bibitem{Q}
B.~Jha, L.~Cueto-Felgueroso, R.~Juanes, Synergetic {Fluid Mixing from Viscous
  Fingering and Alternating Injection}, Phys. Rev. Lett. 111 (2013) 144501.
\newblock \href {https://doi.org/10.1103/PhysRevLett.111.144501}
  {\path{doi:10.1103/PhysRevLett.111.144501}}.

\bibitem{hill1952channeling}
S.~Hill, F.~P, Channeling in packed columns, Chem. Eng. Sci. 1~(6) (1952)
  247--253.
\newblock \href {https://doi.org/https://doi.org/10.1016/0009-2509(52)87017-4}
  {\path{doi:https://doi.org/10.1016/0009-2509(52)87017-4}}.

\bibitem{orr1984use}
F.~M. Orr, J.~J. Taber, Use of {Carbon Dioxide in Enhanced Oil Recovery},
  Science 224~(4649) (1984) 563--569.
\newblock \href {https://doi.org/10.1126/science.224.4649.563}
  {\path{doi:10.1126/science.224.4649.563}}.

\bibitem{perugini2005development}
D.~Perugini, G.~Poli, S.~Rocchi, Development of viscous fingering between mafic
  and felsic magmas: evidence from the {Terra Nova Intrusive Complex
  (Antarctica)}, Mineral Petrol 83~(3-4) (2005) 151--166.
\newblock \href {https://doi.org/10.1007/s00710-004-0064-2}
  {\path{doi:10.1007/s00710-004-0064-2}}.

\bibitem{roy1996patterns}
S.~Roy, S.~Tarafdar, Patterns in the variable {Hele-Shaw} cell for different
  viscosity ratios: {Similarity} to river network geometry, Phys. Rev. E 54
  (1996) 6495--6499.
\newblock \href {https://doi.org/10.1103/PhysRevE.54.6495}
  {\path{doi:10.1103/PhysRevE.54.6495}}.

\bibitem{PINILLA2021e07614}
A.~Pinilla, M.~Asuaje, N.~Ratkovich, Experimental and computational advances on
  the study of {Viscous Fingering: An} umbrella review, Heliyon 7~(7) (2021)
  e07614.
\newblock \href {https://doi.org/10.1016/j.heliyon.2021.e07614}
  {\path{doi:10.1016/j.heliyon.2021.e07614}}.

\bibitem{bischofberger2014fingering}
I.~Bischofberger, R.~Ramachandran, S.~R. Nagel, Fingering versus stability in
  the limit of zero interfacial tension, Nat. Commun. 5~(1) (2014) 1--6.
\newblock \href {https://doi.org/10.1038/ncomms6265}
  {\path{doi:10.1038/ncomms6265}}.

\bibitem{suzuki2020experimental}
R.~X. Suzuki, F.~W. Quah, T.~Ban, M.~Mishra, Y.~Nagatsu, Experimental study of
  miscible viscous fingering with different effective interfacial tension, AIP
  Adv. 10~(11) (2020) 115219.
\newblock \href {https://doi.org/10.1063/5.0030152}
  {\path{doi:10.1063/5.0030152}}.

\bibitem{LI20221598}
D.~Li, Z.~Yang, R.~Zhang, R.~Hu, Y.-F. Chen, Morphological patterns and
  interface instability during withdrawal of liquid-particle mixtures, J.
  Colloid Interface Sci. 608 (2022) 1598--1607.
\newblock \href {https://doi.org/10.1016/j.jcis.2021.10.115}
  {\path{doi:10.1016/j.jcis.2021.10.115}}.

\bibitem{drilling}
X.-B. Huang, J.-S. Sun, Y.~Huang, B.-C. Yan, X.-D. Dong, F.~Liu, R.~Wang,
  {Laponite}: a promising nanomaterial to formulate high-performance
  water-based drilling fluids, Pet. Sci. 18~(2) (2021) 579--590.
\newblock \href {https://doi.org/10.1007/s12182-020-00516-z}
  {\path{doi:10.1007/s12182-020-00516-z}}.

\bibitem{van1977introduction}
H.~Van~Olphen, An {I}ntroduction to {C}lay {C}olloid {C}hemistry: {F}or {C}lay
  {T}echnologists, {G}eologists and {S}oil {S}cientists, Wiley, New York, 1977.

\bibitem{pH}
C.~Martin, F.~Pignon, J.-M. Piau, A.~Magnin, P.~Lindner, B.~Cabane,
  Dissociation of thixotropic clay gels, Phys. Rev. E 66 (2002) 021401.
\newblock \href {https://doi.org/10.1103/PhysRevE.66.021401}
  {\path{doi:10.1103/PhysRevE.66.021401}}.

\bibitem{lagaly2013handbook}
F.~Bergaya, G.~Lagaly, Handbook of {C}lay {S}cience: {P}art {A} {F}undamentals,
  2nd ed., Vol.~5A, Elsevier, Amsterdam, 2013.

\bibitem{bandyopadhyay2004evolution}
R.~Bandyopadhyay, D.~Liang, H.~Yardimci, D.~A. Sessoms, M.~A. Borthwick,
  S.~G.~J. Mochrie, J.~L. Harden, R.~L. Leheny, Evolution of {Particle-Scale
  Dynamics in an Aging Clay Suspension}, Phys. Rev. Lett. 93 (2004) 228302.
\newblock \href {https://doi.org/10.1103/PhysRevLett.93.228302}
  {\path{doi:10.1103/PhysRevLett.93.228302}}.

\bibitem{misra}
C.~Misra, V.~T. Ranganathan, R.~Bandyopadhyay, Influence of medium structure on
  the physicochemical properties of aging colloidal dispersions investigated
  using the synthetic clay {Laponite}\textsuperscript{\textregistered}, Soft
  Matter 17 (2021) 9387--9398.
\newblock \href {https://doi.org/10.1039/D1SM00987G}
  {\path{doi:10.1039/D1SM00987G}}.

\bibitem{delhorme2012monte}
M.~Delhorme, B.~Jönsson, C.~Labbez, Monte {Carlo} simulations of a clay
  inspired model suspension: the role of rim charge, Soft Matter 8 (2012)
  9691--9704.
\newblock \href {https://doi.org/10.1039/C2SM25731A}
  {\path{doi:10.1039/C2SM25731A}}.

\bibitem{ali2016effect}
S.~Ali, R.~Bandyopadhyay, Effect of electrolytes on the microstructure and
  yielding of aqueous dispersions of colloidal clay, Soft Matter 12 (2016)
  414--421.
\newblock \href {https://doi.org/10.1039/C5SM01700A}
  {\path{doi:10.1039/C5SM01700A}}.

\bibitem{Venketesh}
V.~{Thrithamara Ranganathan}, R.~Bandyopadhyay, Effects of aging on the
  yielding behaviour of acid and salt induced {Laponite} gels, Colloids Surf. A
  Physicochem. Eng. Asp. 522 (2017) 304--309.
\newblock \href {https://doi.org/10.1016/j.colsurfa.2017.03.006}
  {\path{doi:10.1016/j.colsurfa.2017.03.006}}.

\bibitem{Shahin2012}
A.~Shahin, Y.~M. Joshi, Physicochemical {Effects in Aging Aqueous Laponite
  Suspensions}, Langmuir 28~(44) (2012) 15674--15686.
\newblock \href {https://doi.org/10.1021/la302544y}
  {\path{doi:10.1021/la302544y}}.

\bibitem{saha2015dynamic}
D.~Saha, R.~Bandyopadhyay, Y.~M. Joshi, {Dynamic Light Scattering Study and
  DLVO Analysis of Physicochemical Interactions in Colloidal Suspensions of
  Charged Disks}, Langmuir 31~(10) (2015) 3012--3020.
\newblock \href {https://doi.org/10.1021/acs.langmuir.5b00291}
  {\path{doi:10.1021/acs.langmuir.5b00291}}.

\bibitem{ruzicka2011fresh}
B.~Ruzicka, E.~Zaccarelli, A fresh look at the {Laponite} phase diagram, Soft
  Matter 7~(4) (2011) 1268--1286.
\newblock \href {https://doi.org/10.1039/C0SM00590H}
  {\path{doi:10.1039/C0SM00590H}}.

\bibitem{Jabbari}
S.~Jabbari-Farouji, H.~Tanaka, G.~H. Wegdam, D.~Bonn, Multiple nonergodic
  disordered states in {Laponite} suspensions: {A} phase diagram, Phys. Rev. E
  78 (2008) 061405.
\newblock \href {https://doi.org/10.1103/PhysRevE.78.061405}
  {\path{doi:10.1103/PhysRevE.78.061405}}.

\bibitem{NICOLAI200151}
T.~Nicolai, S.~Cocard, {Dynamic Light-Scattering Study of Aggregating and
  Gelling Colloidal Disks}, J. Colloid Interface Sci. 244~(1) (2001) 51--57.
\newblock \href {https://doi.org/10.1006/jcis.2001.7930}
  {\path{doi:10.1006/jcis.2001.7930}}.

\bibitem{Angelini2014}
R.~Angelini, E.~Zaccarelli, F.~A. de~Melo~Marques, M.~Sztucki, A.~Fluerasu,
  G.~Ruocco, B.~Ruzicka, Glass--glass transition during aging of a colloidal
  clay, Nat. Commun. 5~(1) (2014) 4049.
\newblock \href {https://doi.org/10.1038/ncomms5049}
  {\path{doi:10.1038/ncomms5049}}.

\bibitem{Anisotropy}
E.~Ben-Jacob, R.~Godbey, N.~D. Goldenfeld, J.~Koplik, H.~Levine, T.~Mueller,
  L.~M. Sander, {Experimental Demonstration of the Role of Anisotropy in
  Interfacial Pattern Formation}, Phys. Rev. Lett. 55 (1985) 1315--1318.
\newblock \href {https://doi.org/10.1103/PhysRevLett.55.1315}
  {\path{doi:10.1103/PhysRevLett.55.1315}}.

\bibitem{nagatsu_matsuda_kato_tada_2007}
Y.~Nagatsu, K.~Matsuda, Y.~Kato, Y.~Tada, Experimental study on miscible
  viscous fingering involving viscosity changes induced by variations in
  chemical species concentrations due to chemical reactions, J. Fluid Mech. 571
  (2007) 475–493.
\newblock \href {https://doi.org/10.1017/S0022112006003636}
  {\path{doi:10.1017/S0022112006003636}}.

\bibitem{hirata1998fracturing}
T.~Hirata, Fracturing due to fluid intrusion into viscoelastic materials, Phys.
  Rev. E 57 (1998) 1772--1779.
\newblock \href {https://doi.org/10.1103/PhysRevE.57.1772}
  {\path{doi:10.1103/PhysRevE.57.1772}}.

\bibitem{paterson1981radial}
L.~Paterson, Radial fingering in a {Hele Shaw} cell, J. Fluid Mech. 113 (1981)
  513--529.
\newblock \href {https://doi.org/10.1017/S0022112081003613}
  {\path{doi:10.1017/S0022112081003613}}.

\bibitem{lajeunesse19973d}
E.~Lajeunesse, J.~Martin, N.~Rakotomalala, D.~Salin, 3{D} instability of
  miscible displacements in a {Hele-Shaw} cell, Phys. Rev. Lett. 79 (1997)
  5254--5257.
\newblock \href {https://doi.org/10.1103/PhysRevLett.79.5254}
  {\path{doi:10.1103/PhysRevLett.79.5254}}.

\bibitem{van1986fractal}
H.~Van~Damme, F.~Obrecht, P.~Levitz, L.~Gatineau, C.~Laroche, Fractal viscous
  fingering in clay slurries, Nature 320~(6064) (1986) 731--733.
\newblock \href {https://doi.org/10.1038/320731a0}
  {\path{doi:10.1038/320731a0}}.

\bibitem{VanDamme1990}
H.~V. Damme, E.~Lemaire, From {F}low to {F}racture and {F}ragmentation in
  {C}olloidal {M}edia, {I}n: {J.C.} {C}harmet, {S}. {R}oux, {E}. {G}uyon
  ({E}ds.), {D}isorder and {F}racture., Springer, Boston, MA, 1990, pp.
  83--104.
\newblock \href {https://doi.org/10.1007/978-1-4615-6864-3_6}
  {\path{doi:10.1007/978-1-4615-6864-3_6}}.

\bibitem{Bittleston2002}
S.~H. Bittleston, J.~Ferguson, I.~A. Frigaard, Mud removal and cement placement
  during primary cementing of an oil well -- {L}aminar non-{N}ewtonian
  displacements in an eccentric annular {Hele-Shaw} cell, J. Eng. Math. 43~(2)
  (2002) 229--253.
\newblock \href {https://doi.org/10.1023/A:1020370417367}
  {\path{doi:10.1023/A:1020370417367}}.

\bibitem{battery}
W.~Xu, J.~Wang, F.~Ding, X.~Chen, E.~Nasybulin, Y.~Zhang, J.-G. Zhang, Lithium
  metal anodes for rechargeable batteries, Energy Environ. Sci. 7 (2014)
  513--537.
\newblock \href {https://doi.org/10.1039/C3EE40795K}
  {\path{doi:10.1039/C3EE40795K}}.

\bibitem{gorell1983theory}
S.~B. Gorell, G.~Homsy, A theory of the optimal policy of oil recovery by
  secondary displacement processes, SIAM J. Appl. Math. 43~(1) (1983) 79--98.
\newblock \href {https://doi.org/10.1137/0143007} {\path{doi:10.1137/0143007}}.

\bibitem{Gao2019}
T.~Gao, M.~Mirzadeh, P.~Bai, K.~M. Conforti, M.~Z. Bazant, Active control of
  viscous fingering using electric fields, Nat. Commun. 10~(1) (2019) 4002.
\newblock \href {https://doi.org/10.1038/s41467-019-11939-7}
  {\path{doi:10.1038/s41467-019-11939-7}}.

\bibitem{Li}
S.~Li, J.~S. Lowengrub, J.~Fontana, P.~Palffy-Muhoray, {Control of Viscous
  Fingering Patterns in a Radial Hele-Shaw Cell}, Phys. Rev. Lett. 102 (2009)
  174501.
\newblock \href {https://doi.org/10.1103/PhysRevLett.102.174501}
  {\path{doi:10.1103/PhysRevLett.102.174501}}.

\bibitem{Zheng}
Z.~Zheng, H.~Kim, H.~A. Stone, {Controlling Viscous Fingering Using
  Time-Dependent Strategies}, Phys. Rev. Lett. 115 (2015) 174501.
\newblock \href {https://doi.org/10.1103/PhysRevLett.115.174501}
  {\path{doi:10.1103/PhysRevLett.115.174501}}.

\bibitem{borgman2019immiscible}
O.~Borgman, T.~Darwent, E.~Segre, L.~Goehring, R.~Holtzman, Immiscible fluid
  displacement in porous media with spatially correlated particle sizes, Adv.
  Water Resour. 128 (2019) 158--167.
\newblock \href {https://doi.org/10.1016/j.advwatres.2019.04.015}
  {\path{doi:10.1016/j.advwatres.2019.04.015}}.

\bibitem{glucose}
C.~Talon, L.~J. Smith, J.~W. Brady, B.~A. Lewis, J.~R. Copley, D.~L. Price,
  M.-L. Saboungi, {Dynamics of Water Molecules in Glucose Solutions}, J. Phys.
  Chem. B 108~(16) (2004) 5120--5126.
\newblock \href {https://doi.org/10.1021/jp035161e}
  {\path{doi:10.1021/jp035161e}}.

\bibitem{dmf}
Y.~Lei, H.~Li, H.~Pan, S.~Han, {Structures and Hydrogen Bonding Analysis of
  N,N-Dimethylformamide and N,N-Dimethylformamide -- Water Mixtures by
  Molecular Dynamics Simulations}, J. Phys. Chem. A 107~(10) (2003) 1574--1583.
\newblock \href {https://doi.org/10.1021/jp026638+}
  {\path{doi:10.1021/jp026638+}}.

\end{thebibliography}
	
	\pagebreak
	\renewcommand{\figurename}{Supplementary Fig.}
	\begin{center}
		\textbf{\LARGE Supplementary Material}\\
		\vspace{0.4cm}
		\textbf{\LARGE \textcolor{blue}{Pattern selection in radial displacements of a confined aging viscoelastic fluid}}\\
		\vspace{0.4cm}
		{Palak}$^\dagger$, {Vaibhav Raj Singh Parmar}$^\ddagger$, {Debasish Saha}$^\S$, and {Ranjini Bandyopadhyay}$^*$\\
		\vspace{0.4cm}
		\textit{Soft Condensed Matter Group, Raman Research Institute, C. V. Raman Avenue, Sadashivanagar, Bangalore 560 080, INDIA}
		
		\date{\today}
	\end{center}
	
	\footnotetext[2]{palak@rri.res.in}
	\footnotetext[3]{vaibhav@rri.res.in}
	\footnotetext[4]{Debasish.Saha@uni-duesseldorf.de}
	\footnotetext[1]{Corresponding Author: Ranjini Bandyopadhyay; Email: ranjini@rri.res.in}
	\maketitle
	
	\definecolor{black}{rgb}{0.0, 0.0, 0.0}
	\definecolor{red(ryb)}{rgb}{1.0, 0.15, 0.07}
	\definecolor{darkred}{rgb}{0.55, 0.0, 0.0}
	\definecolor{blue(ryb)}{rgb}{0.01, 0.2, 1.0}
	\definecolor{darkcyan}{rgb}{0.0, 0.55, 0.55}
	\definecolor{navyblue}{rgb}{0.0, 0.0, 0.5}
	\definecolor{olivedrab(web)(olivedrab3)}{rgb}{0.42, 0.56, 0.14}
	\definecolor{darkraspberry}{rgb}{0.50, 0.0, 1.0}
	\definecolor{magenta}{rgb}{1.0, 0.0, 1.0}
	\newcommand{\pcircle}{\textcolor{darkraspberry}{\large$\bullet$}}
	\newcommand{\phex}{\textcolor{darkraspberry}{\large$\varhexagonblack$}}
	\newcommand{\rdhexagon}{\textcolor{red(ryb)}{\small$\blackdiamond$}}

	\setcounter{table}{0}
	\renewcommand{\thetable}{S\arabic{table}}%
	\renewcommand{\tablename}{Supplementary Table}
	\setcounter{figure}{0}
	\makeatletter 
	\renewcommand{\figurename}{Supplementary Fig.}
	\setcounter{figure}{0}
	\makeatletter 
	\renewcommand{\thefigure}{S\arabic{figure}}
	\setcounter{section}{0}
	\renewcommand{\thesection}{ST\arabic{section}}
	\setcounter{equation}{0}
	\renewcommand{\theequation}{S\arabic{equation}}
	
	\section{Additional experimental details}
	\subsection{\label{em:ssp} Sample preparation} 
	Hygroscopic Laponite\textsuperscript{\textregistered} XLG powder (BYK additives Inc.) was baked in an oven at 120$^{\circ}$C for 18-20 hours prior to sample preparation. We prepared an aqueous suspension of concentration 3.25\% w/v by adding 1.625 g of dried Laponite powder to 50~ml double distilled water (Millipore Corp., resistivity 18.2 MΩ.cm). We stirred the sample continuously and vigorously for 45 minutes at room temperature (25$^{\circ}$C), and the resulting suspension was filtered through a syringe filter (Millex\textsuperscript{\textregistered}, Sigma Aldrich) of pore size 0.45~$\mu$m.
	\subsection{\label{em:sp} Hele-Shaw Cell}
	A radial Hele-Shaw cell (Fig.~2(a) in the main paper) comprising two circular glass plates, each of radius 30~cm and thickness 10 mm, was used to study interfacial patterns. Teflon spacers were used to ensure a constant gap of $170~\mu m$ between the plates. An aqueous clay suspension, the displaced fluid in our experiments, was first filled in this quasi two-dimensional geometry with a syringe pump (NE-8000, New Era Pump Systems, USA) at a constant volumetric flow rate of 20~ml/min through a 3~mm hole drilled in the top plate. Water and mineral oil were used as the displacing fluids in two separate sets of experiments and were injected through the same hole after a pre-determined time $t_w$ at flow rates ($q$) ranging from 1~ml/min to 50~ml/min. We used water and mineral oil for the miscible (interfacial tension $\sigma = 0$) and immiscible ($\sigma \neq 0$) displacement experiments, respectively. Water and mineral oil were respectively dyed with Rhodamine B ($\geq$95\% (HPLC)) and Oil Red O, procured from Sigma Aldrich, to ensure adequate contrast at the interface. We recorded the growth of the interfacial patterns with a DSLR camera (D5200, Nikon, Japan) that was set up below the Hele-Shaw cell with a spatial resolution of 1920$\times$1080 pixels (one pixel area = $2.2 \times 10^{-3}$ cm$^2$) and a frame rate of 30~fps. An area enclosed by a circle of radius 1.18 cm is excluded to avoid experimental artifacts in image analysis. 
	\subsection{\label{em:spRheology} Rheology}
	The sample preparation protocol for the rheological measurements was identical to that for the Hele-Shaw experiments. A stress controlled Anton Paar MCR 501 rheometer was used for all rheology measurements reported here. For these measurements, the sample temperature was maintained at room temperature (25$^{\circ}$C) using a Peltier temperature control device (C/PPTD 180/MD). We loaded the freshly prepared Laponite suspension in the cone and plate geometry and shear melted it at a high shear rate of 500~$\mathrm{s^{-1}}$ to remove any memory of the sample loading process. The time at which shear application was stopped and the sample was allowed to evolve spontaneously in the rheological experiments was noted as aging time $t_w$ = 0. The elastic and viscous moduli ($G^{\prime}$ and $G^{\prime\prime}$) of Laponite suspensions as a function of $t_w$ were recorded at an applied strain amplitude 0.5 \% and angular frequency 1 rad/s.
	
	\begin{figure}[H]
		\centering
		\includegraphics[width=6.5in, keepaspectratio]{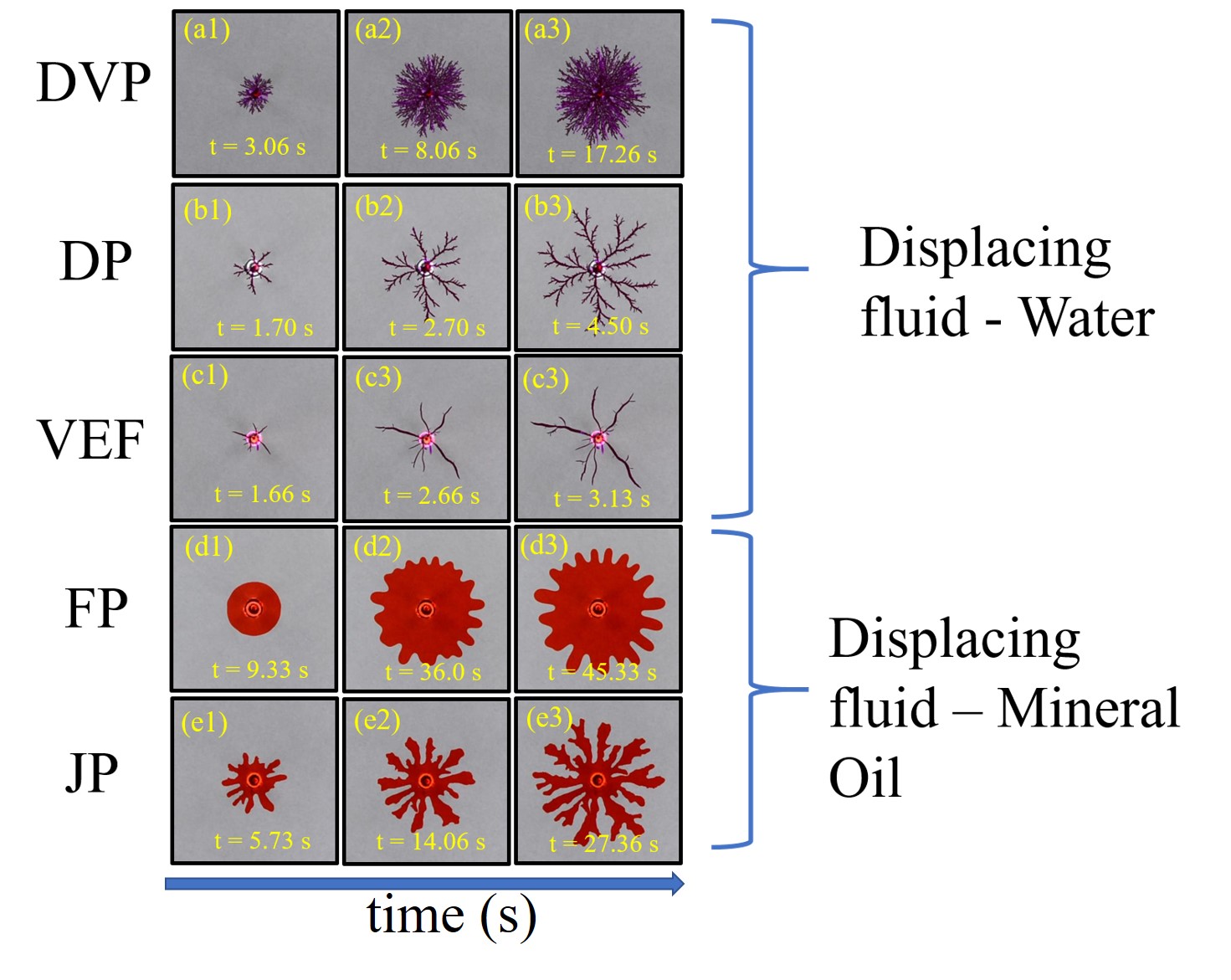}
		\caption{Raw RGB images illustrating distinct morphologies of interfacial patterns obtained during the miscible and immiscible displacements of aging Laponite\textsuperscript{\textregistered} suspensions at constant flow rate $q$ = 5 ml/min. (a1-a3) Temporal evolution of dense viscous pattern (DVP) at $t_w$ = 1 h for an interfacial tension $\sigma$ = 0. (b1-b3) Temporal evolution of dendritic pattern (DP) for $t_w$ = 3 h and $\sigma$ = 0. (c1-c3) Temporal evolution of viscoelastic fracture (VEF) at $t_w$ = 24 h and $\sigma$ = 0. (d1-d3) Temporal evolution of flower pattern (FP) at $t_w$ = 1 h and $\sigma$ $\ne$ 0. (e1-e3) Temporal evolution of jagged pattern (JP) at $t_w$ = 24 h and $\sigma$ $\ne$ 0.}
		\label{sizeratio-EP}
	\end{figure}
	
	\begin{figure}[H]
		\centering
		\includegraphics[width=5.0in]{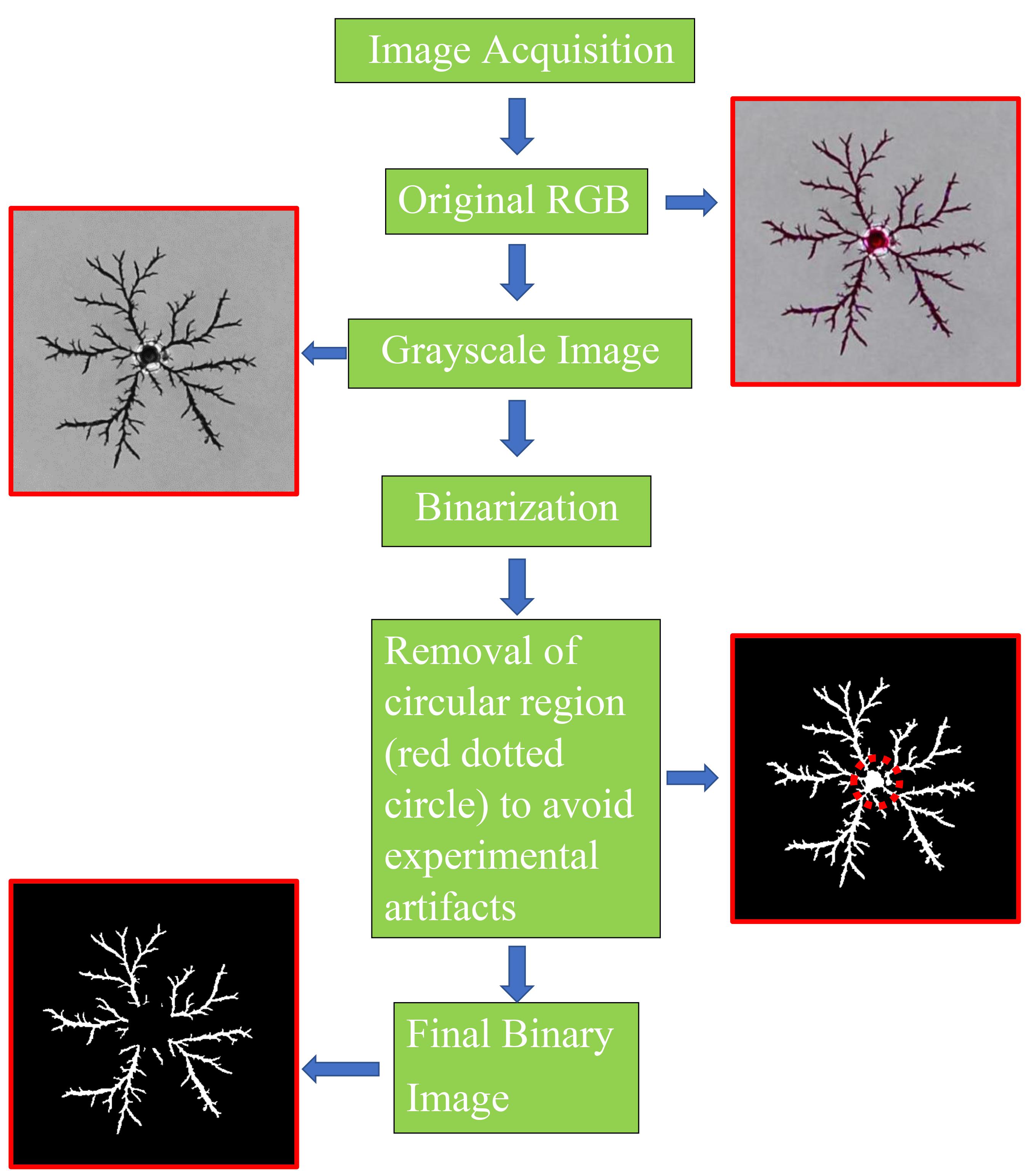}
		\caption{Flow chart illustrating the binarization of RGB images.}
		\label{binary-EP} 
	\end{figure}

	\begin{figure}[!t]
		\centering
		\includegraphics[width=6.0in]{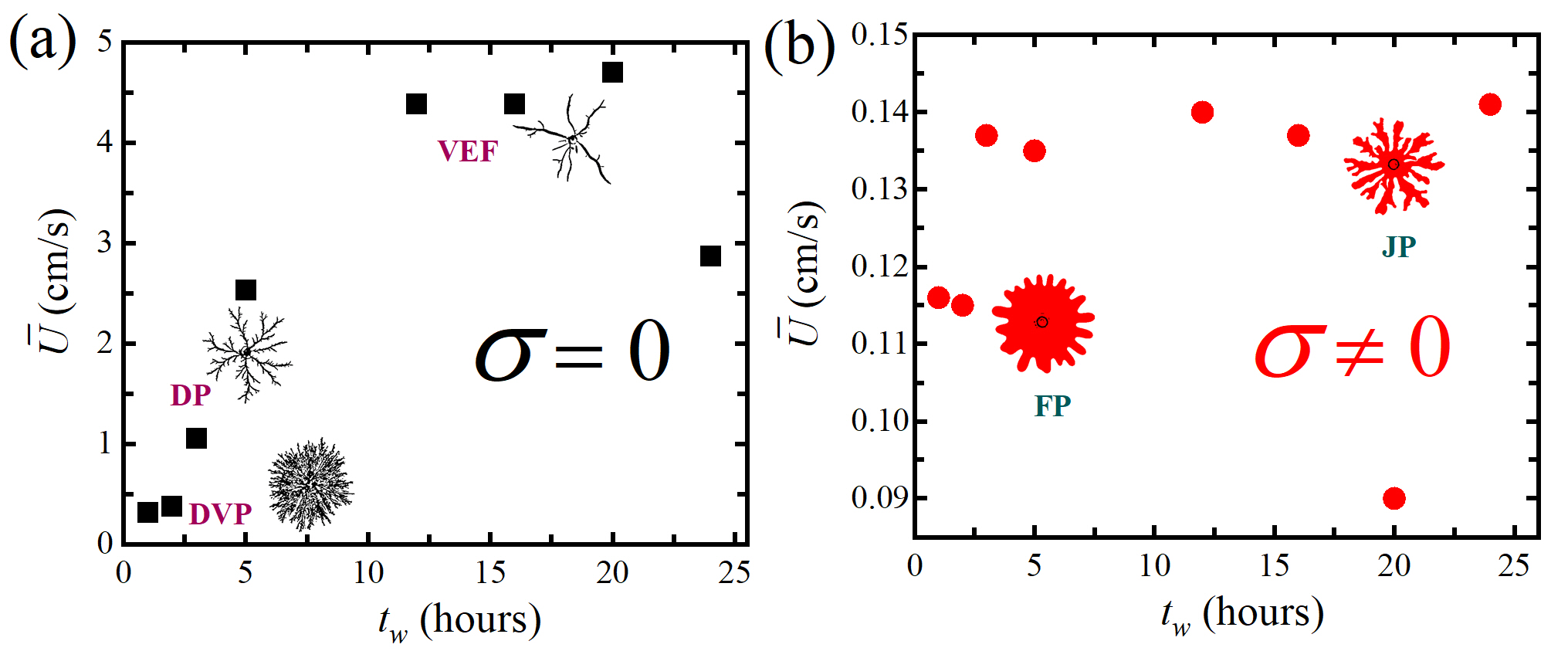}
		\caption{The time-averaged finger-tip velocities vs. aging time $t_w$ for (a) miscible (\blsquare) and (b) immiscible (\rcircle) displacements of Laponite suspensions by Newtonian fluids (water and oil respectively) at constant flow rate $q$ = 5 ml/min. DVP, DP, VEF, FP and JP refer to the experimentally observed dense viscous patterns, dendritic patterns, viscoelastic fractures, flower patterns and jagged patterns, respectively.}
		\label{tw-EP} 
	\end{figure}
	
	\begin{figure}[H]
		\centering
		\includegraphics[width=2.5in]{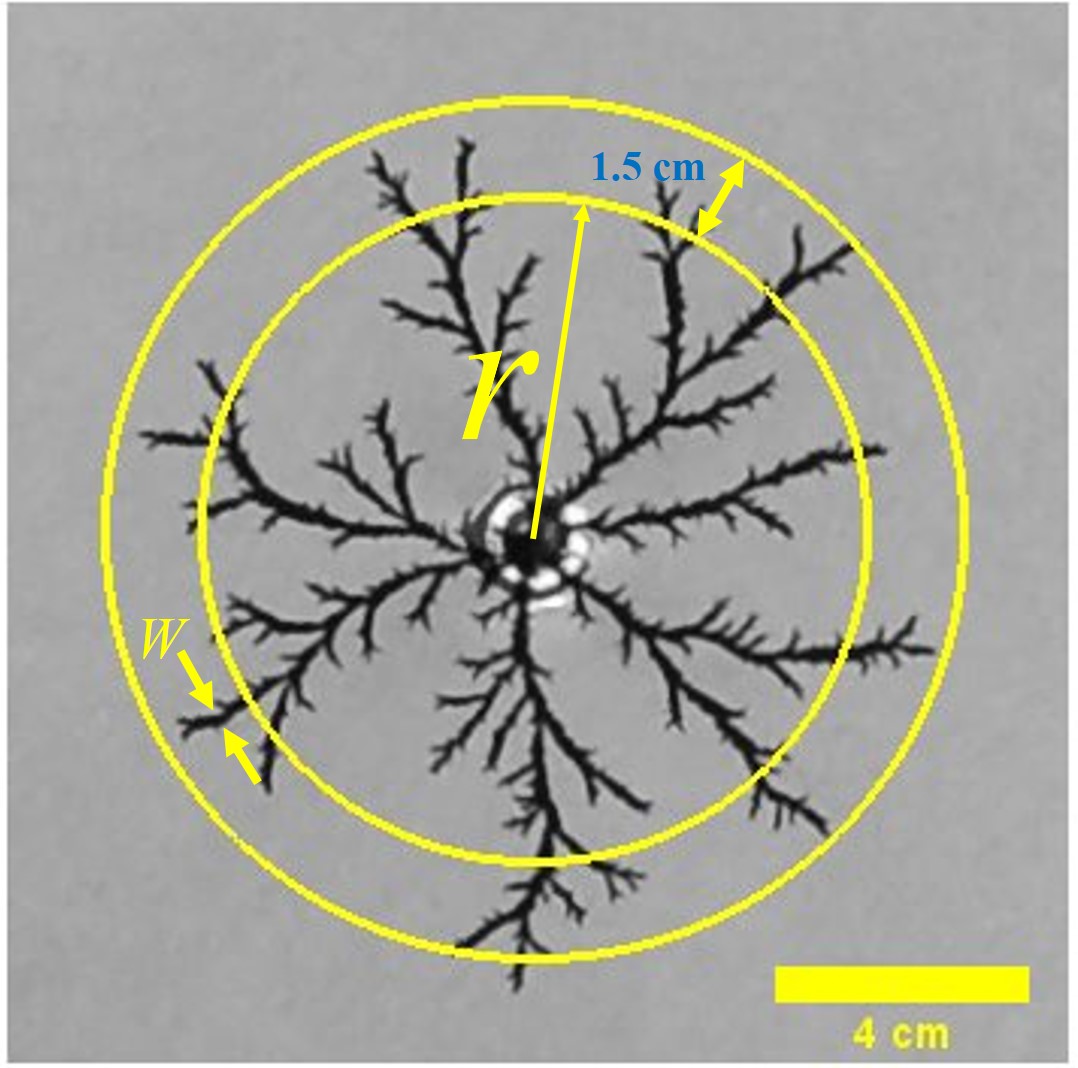}
		\caption{A representative interfacial pattern showing the annular region of width 1.5 cm, used in the calculations of average branch frequencies ($f$) and branch widths ($w$). The average branch frequency, $f$, is estimated as $f$ = $\frac{1}{m}$ $\sum_{j = 1}^{m} \frac{N_j}{2 \pi r_j}$ over $m = 32$ circles within the annular region. Here, $N_j$ is the number of branches intersecting a circle of radius $r_j$, such that 5.2 cm $\leq r_j \leq$ 6.7 cm. The average branch width, $w$, of an interfacial patterns is estimated as $w$ = $\frac{1}{m}$ $\sum_{j = 1}^{m} <W>_{r_j}$ where $<W>_{r_j}$ denotes the average width of the branches intersecting a circle of radius $r_j$ in the same annular region of interest. }
		\label{inset-EP} 
	\end{figure}

	\section{Calculations of viscosity ratios}
\begin{figure}[!h]
		\includegraphics[width=6.5in ]{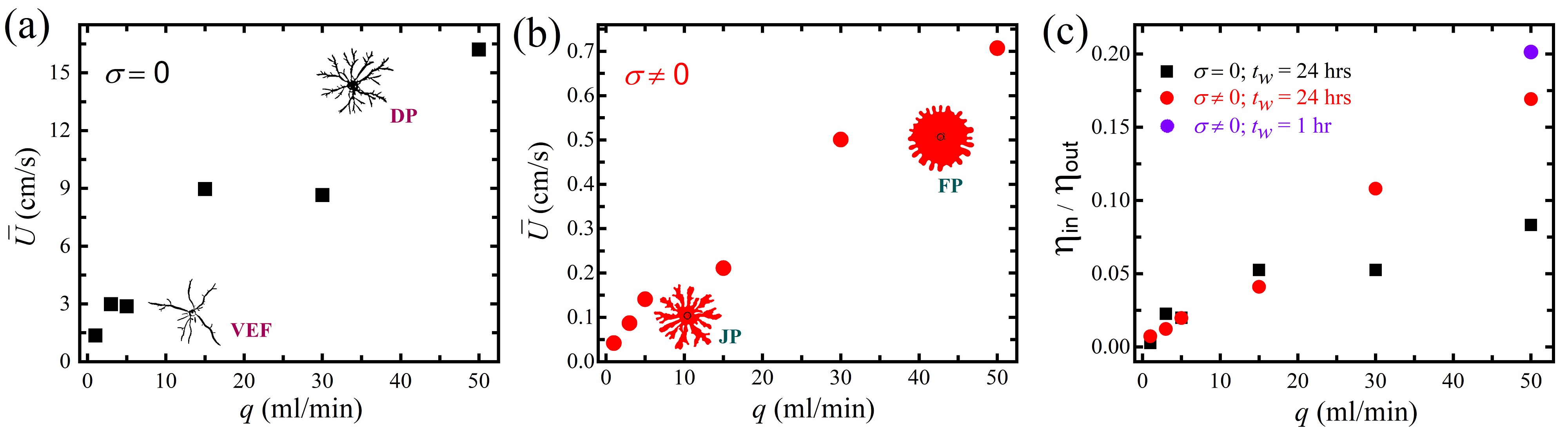}
		\centering
		\caption{The time-averaged finger-tip velocities $\bar{U}$ vs. $q$ for (a) miscible (\blsquare) and (b) immiscible (\rcircle) displacements of Laponite suspensions at age $t_w$ = 24 hrs. (c) Viscosity ratios $\eta_{in}/\eta_{out}$ vs. $q$, where $\eta_{in}$ is the viscosity of the displacing Newtonian fluids and $\eta_{out}$ is the shear-dependent viscosity of the displaced Laponite suspension. The viscosity ratio corresponding to a stable pattern observed during displacement of a Laponite suspension of $t_w$ = 1 h by mineral oil at $q$ = 50 ml/min (\pcircle) is also included.}
		\label{q-EP} 
	\end{figure}

	The values of the time-averaged finger-tip velocity $\bar{U}$ and shear rates $\dot{\gamma}$ are estimated using the protocols mentioned in section 3.1 in the main paper. For increase in the injection flow rate, $q$, of the displacing fluid, the estimated values of $\bar{U}$ vary in the range of $1.35-16.21~\mathrm{cm/s}$ for miscible (Supplementary Fig.~\ref{q-EP}(a)) and $0.04-0.70~\mathrm{cm/s}$ for immiscible displacements (Supplementary Fig.~\ref{q-EP}(b)). The shear-dependent viscosities $\eta_{out}$ of the displaced Laponite suspensions are extracted  at shear rates ($2\bar{U}/b$) from the data in Fig.~1(b). The viscosities $\eta_{in}$ of the displacing Newtonian fluids are extracted at shear rate $\dot{\gamma}\rightarrow0$ from the data in Fig.~1(c). The viscosity ratios $\eta_{in}/\eta_{out}$ of the fluid pairs are then evaluated and plotted in Fig.~\ref{q-EP}(c).

	\begin{figure}[!b]
		\centering
		\includegraphics[width=4.0in]{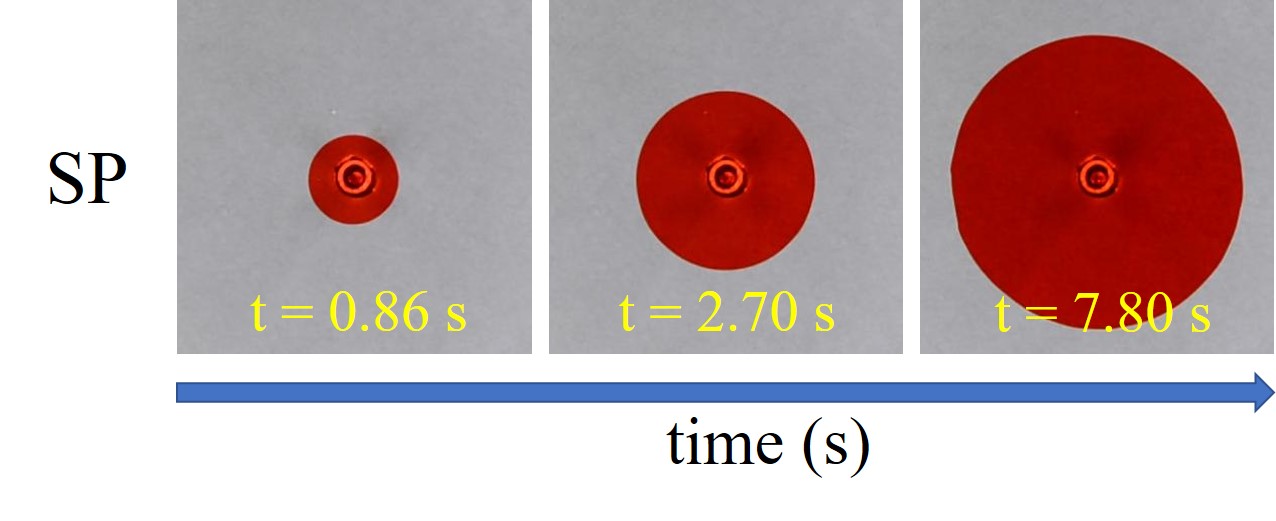}
		\caption{Temporal evolution of stable patterns (SP) at $t_w$ = 1 h, $q$ = 50 ml/min and $\sigma$ $\ne$ 0 in RGB format.}
		\label{SP-EP} 
	\end{figure}
	
\end{document}